\documentclass[preprint,12pt,3p]{article} 

\usepackage{soul}
\usepackage{color}

\usepackage{graphicx}

\usepackage{amssymb}

\usepackage{subfigure}
\graphicspath{{figure/}}

\usepackage{epstopdf}

\title{Graphene Healing Mechanisms: A Theoretical Investigation} 
\author{Tiago Botari\footnote{Applied Physics Department, State University of Campinas (UNICAMP), 13083-859 Campinas-SP, Brazil.} \footnote{Tiago Botari: tiagobotari@gmail.com; D.S. Galvão: galvao@ifi.unicamp.br}, Ricardo Paupitz\footnote{Physics Department, Univ. Estadual Paulista (UNESP), 13506-900 Rio Claro-SP, Brazil.},\\ Pedro Alves da Silva Autreto$^*$, Douglas S. Galvao$^*$}

\begin{document}


\maketitle


\begin{abstract}
Large holes in graphene membranes were recently shown to heal, either at room temperature during a low energy STEM experiment, or by annealing at high temperatures. However, the details of the healing mechanism remain unclear.
We carried out fully atomistic reactive molecular dynamics simulations in order to address these mechanisms under different experimental conditions.
Our results show that, if a carbon atom source is present, high temperatures can provide enough energy for the carbon atoms to overcome the potential energy barrier and to produce perfect reconstruction of the graphene hexagonal structure. At room temperature, this perfect healing is only possible if the heat effects of the electron beam from STEM experiment are explicitly taken into account.
The reconstruction process of a perfect or near perfect graphene structure involves the formation of linear carbon chains, as well as rings containing 5, 6, 7 and 8 atoms with planar ({\sl Stone-Wales} like) and non-planar ({\sl lump} like) structures.  These results shed light on the healing mechanism of graphene when subjected to different experimental conditions. Additionally, the methodology presented here can be useful for investigating the tailoring and manipulations of other nano-structures.
\end{abstract}



\section{Introduction} 

Graphene, a two-dimensional carbon allotrope, has unique electronic, thermal and mechanical properties \cite{Geim2007}.
Originally, this material was obtained from graphite using an exfoliation process, called the ``scotch tape'' method \cite{Novoselov2004}. Although this method yields pristine graphene samples, such a process is costly and not easily scalable. Presently, chemical vapor deposition (CVD) has been the most widely used process to grow graphene samples on a diverse set of substrates, such as steel \cite{Yuan2009}, Ni \cite{Obraztsov2007,Reina2009}, and Cu \cite{Niu2013}.

Graphene properties are extremely sensitive even to small modifications in its honeycomb structure \cite{Banhart2011}.  Inherent defects from CVD growth processes represent obstacles to some technological applications, because they can degrade graphene electronic and mechanical properties. 
On the other hand, defects can be usefully exploited to obtain different properties for specific applications \cite{Vicarelli}. 

Recently, it was demonstrated that etched nanoholes of up to 100 vacancies on graphene membranes can be healed under low power STEM observation, even at room temperature \cite{Zan2012}. The healing effect consists of the reconstruction (knitting) of the graphene structure.  Carbon atoms from external sources near the hole region, eventually interact with its edges and can fill the vacancies. Hydrocarbon impurities near the membrane can also serve as a source for these extra carbon atoms. These nanohole fillings can occur with the formation of either non-hexagonal, near-amorphous, or perfectly hexagonal structures \cite{Zan2012}. 

Other experimental works have also addressed the reconstruction of mono-vacancies and holes in graphene. Chen {\it et al.} \cite{Chen2013} demonstrated the effectiveness of thermal annealing up to $900~^o$C ($1173.15~$K) to heal defects in graphene membranes. Kholmonov {\it et al.} \cite{Kholmanov2011} demonstrated defect healing in the top layer of multilayer graphene {\it via} CVD techniques using acetylene as a carbon feedstock and iron (Fe) as a catalyst at $900~^o$C ($1173.15~$K). The evolution and control of nanoholes in graphene by carbon atom thermal-induced migrations were studied by Xu {\it et al.} \cite{Xu2012}. For temperatures around $525~^o$C ($798.15$K), graphene oxide (GO) can be healed and simultaneously reduced by methane plasma resulting in high quality graphene \cite{Cheng2012}. Moreover, self-repair mechanism during graphene sculpting by a focused electron beam were observed by Song {\it et al.} \cite{Song2011}, while transformation of  amorphous carbon into graphene was investigated by Barreiro {\it et al.} \cite{Barreiro2013}. 
 
To address these healing and self-repair mechanisms, we carried out fully atomistic molecular dynamics simulations using the ReaxFF reactive force field\cite{vanDuin2001}. Initially, we investigated the potential energy landscape near a graphene nanohole (defective regions) using a carbon atom as probe to estimate the energies involved during the healing process. Then, we carried out simulations at different temperatures (controlled by thermostats) to investigate the role of thermal energy in these healing mechanisms. Finally, we simulated the heating effects of an electron beam scanning to address the so called self-healing mechanism at room temperature reported by Zan {\it et al.} \cite{Zan2012}.
Our simulations show that graphene healing can be obtained by a simple annealing at high temperature in the presence of a carbon source. However, we could not observe the holes being filled at room temperature because of the energy barriers involved in the process.  
But if we take into account the electron beam heating effects in the simulations, the graphene healing at room temperature can be observed. In this case, either an imperfect or even a perfect hexagonal structure, depending on the specific energy rate for the beam heating, are obtained.

\section{Methods and Model}

We carried out fully atomistic reactive molecular dynamics simulations (MD) using the ReaxFF force field  \cite{vanDuin2001}, as implemented in the Large-scale atomic/ molecular massively parallel simulator (LAMMPS) package \cite{Plimpton19951,lammps-reaxff}. ReaxFF is a reactive force field that allows the study of formation and dissociation of chemical bonds with lower computational cost in comparison to {\it{ab initio}} methods. ReaxFF parametrization is based on density functional theory (DFT) calculations and was successfully used to investigate many dynamical and chemical processes \cite{Botari2014,ISI:000310460300017,fluoro2013}. In the present work, we used the Chenoweth et al. (2008) C/H/O ReaxFF parametrization\cite{ISI:000252815100034} and simulations were carried out with a time steps of $0.1~fs$, with temperatures controlled through a Nos\'e-Hoover thermostat \cite{Nosehoover,Shuichi01011991}.

For comparison with ReaxFF results, we have also carried out some calculations using the Self-Consistent Charge Density Functional Tight-Binding (SCC-DFTB) \cite{DFTB,SCCDFTB} method, as implemented on DFTB+ package \cite{DFTBplus}. DFTB is a DFT-based method and can handle large systems. SCC-DFTB is an implementation of DFTB approach that has the advantage of using Mulliken self-consistent charge redistribution (SCC), which corrects some deficiencies of standard DFTB methods \cite{SCCDFTB}. In general, dispersion terms are not considered in DFTB methods and were included here via Slater-Kirkwood Polarizable atomic model \cite{DFTBplus}.

\begin{figure}
\centering
\includegraphics[scale=0.28]{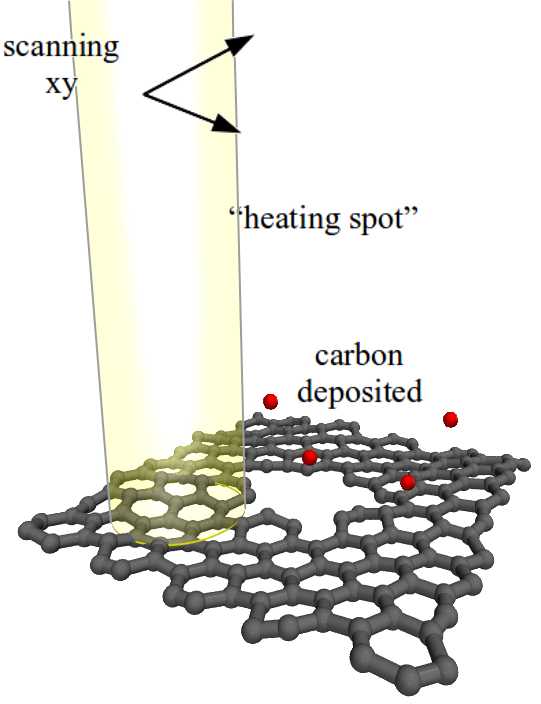}
\caption{Schematic representation of the computational model considered in our simulations. Local heating 
is represented by the yellow region, while added atoms (carbon atoms deposited) are represented in red.}
\label{figure1}
\end{figure}

The computational model used in our calculations consisted of a single-layer graphene membrane (aligned along the $xy$ plane) with a hole ($3.2$ \AA~ of radius) in its center. We considered two simulation scenarios: (i) the healing mechanism dependence on temperature and; (ii) mimicking effects of an electron beam scanning to trigger the healing mechanism at room temperature. Scenario (i) was implemented using temperatures ranging from $300$K up to $2000$K. For scenario (ii), in order to simulate the heat effect induced by the electron beam interaction with the system \cite{Ramasse2012,Krasheninnikov2007,Krasheninnikov2010, Banhart,Petkov2013}, we have applied a local heating protocol in which a rescaling of atomic velocities inside a cylindrical region is performed (see figure \ref{figure1}). The position of the heated region was varied during the simulations, thus mimicking the STEM experiments. 
Also, for both scenarios, we restricted the movement of atoms located at the edges of the graphene membrane by using virtual springs with elastic constants $K=~30.0~K$cal/mol.\AA. For scenario (ii) we also fixed the temperature in these atoms in order to dissipate the accumulation of energy. In our simulations, additional carbon atoms (called ``added atoms'' in the text and colored in red in all figures) with random kinetic energy values were deposited at random positions and at regular intervals of $500~ fs$. 

Carbon depositions were made using single atoms, but we need to emphasise that depending on specific experimental conditions, several small hydrocarbons or other carbon species ($C_2$ and $C_3$) can be present. For instance, in a methane plasma \cite{Jacob1994} it is expected the presence of  CH, CH2 and CH3 species in the environment \cite{Davies1992}. However, in order to focus the investigation of the healing mechanism and speed up the molecular dynamics simulations, we opted to use only single carbon atoms for the deposition processes. The inclusion of other hydrocarbons, it would make necessary to consider other complicated questions, such as the activation energies and heat of formation for each considered species. These questions, despite of being interesting, would bring unnecessary complications to our analysis.

The local heating protocol adds a non-translational kinetic energy to the atoms in a cylindrical region centered along the $z$ axis perpendicular to the graphene membrane (figure \ref{figure1}). The local heating scanned the membrane along the $x$ and $y$ directions, similarly to what is done by an electron beam in a STEM experiment. The effects of local heating with a cylindrical radius of $3.5$ \AA~ and energy rates between $0-3.0~ kcal/(mol.fs)$ were investigated. If the energy rate per area is maintained constant, we can expect that changes in the cylindrical radius would not change the necessary energy rate to trigger the healing process. On the other hand, we expect that a small cylindrical radius would increase the necessary time of scanning to obtain a complete healing.
All results presented throughout the text are representative simulations that illustrate typical results for the different processes of the graphene healing mechanisms.

\section{Results and Discussion}

\subsection{Potential Energy Landscape}

First, we will discuss the potential energy experienced by a carbon atom probe placed near to a defective graphene structure containing a large hole, {\it i.e.}, a energy landscape mapping.
These mappings were generated for the $xy$ plane while considering different fixed out-of-plane probe distances, {\it i.e.}, different $z$-coordinates values measured from the graphene basal plane reference (see figure 1). 
From these mappings, it is possible to identify the most reactive and repulsive regions experienced by the probe atom. Through this analysis we can obtain a reasonable evaluation for the threshold energy values involved in the healing processes.

\begin{figure*}[t]
\centering
\includegraphics[width=1.0\linewidth]{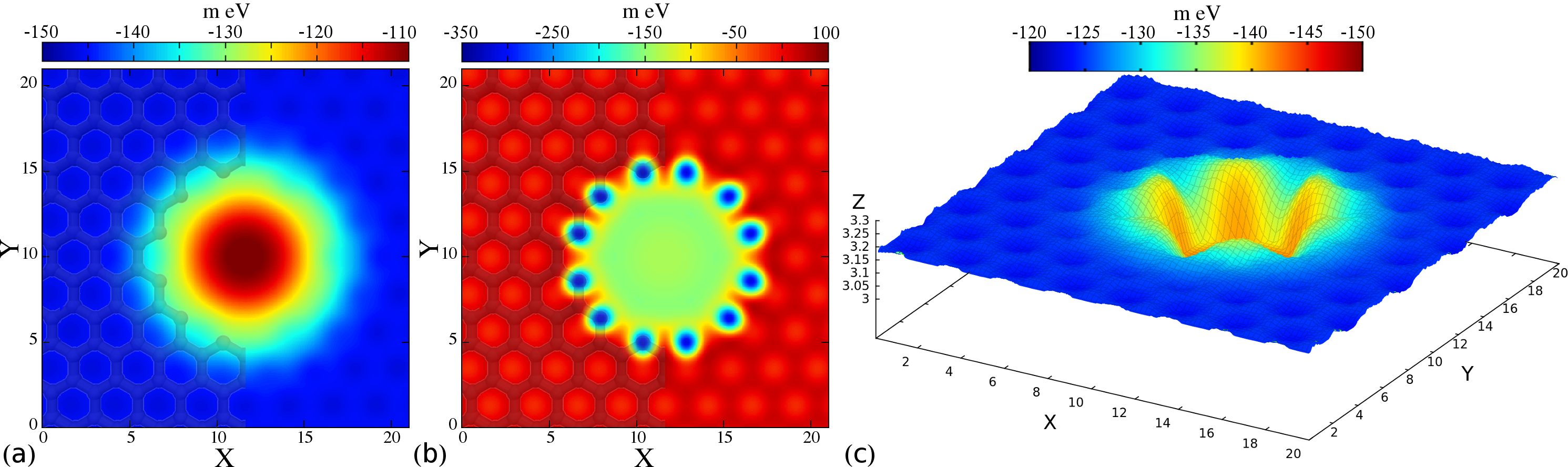}
\caption{Potential energy (landscape energy) for a probe atom placed above the graphene membrane with a hole of radius $4.5$ \AA; (a) potential energy for probe atom at a distance of $3.2$ \AA~ from graphene; (b) potential energy for a probe atom placed at $2.2$ \AA~; and (c) Minimum potential energy map for a probe atom near a hole of radius $3.2$ \AA~ etched in a graphene membrane. For each ($x$,$y$) position we represent the height ($z$ direction) position in which the free atom potential energy reaches a minimum, while the color represents the value of this potential energy for that specific point.}
\label{figure2}
\end{figure*}

For a free atom on the membrane surface ({\it i.e.}, an atom not covalently bonded to graphene), the equilibrium distance is around $3.2$ \AA. 
At this distance, it is energetically favorable for an atom to be placed above the hollow site of a hexagonal ring rather than above a carbon atom.
In the case of a hole etched on the membrane, the probe atom starts to be repelled as it approaches the 
hole borders, because of the decrease on the van der Waals interaction between the atom and the membrane. This interpretation is supported by results shown in figure \ref{figure2} (a). 
If we consider, for instance, that a free atom is at a $z$ distance $2.2 $ \AA~ above the membrane, the 
lattice becomes highly repulsive and a pronounced minimum energy takes place near the hole edges indicating highly chemical reactive regions, as shown 
in figure \ref{figure2}(b). 

Depending on the $z$ distance between the free atom and the membrane surface, repulsive or attractive regions exist. 
Therefore, to start filling the hole with free atoms on the membrane surface, the free atoms must be able to easily migrate to the defective regions, {\it i.e.}, these free atoms must have sufficient energy to overcome the energetic barrier found at the hole edges. These barrier energies can be visualized by mapping the minimum landscape potential energy over the $z$ distance for each planar coordinate of the probe atom, as shown in figure \ref{figure2} (c). The energetic barrier near the edge of the hole as well as the potential energy inside the hole region is significantly higher than that on the surface, as highlighted in reddish colors. However, if the free atom reaches the empty region (inside the hole), there is a high probability for it to be trapped. As soon as the atom is trapped the reaction with the edge of the hole is facilitated. Other potential energy landscapes for different hole sizes can be found in supplementary material.\cite{supplementary_material}

The energy mapping profiles indicate that the healing mechanism is not effective at room temperature because the probability for a free atom to overcome the energetic barrier and enter into the hole region is quite low. In fact, for a distance of $3.2 $ \AA~ from the membrane, the difference in the potential energy between the hole center and a hollow site has a value around $40~ meV$ considering a hole with a radius of $4.5$\AA~, while the thermal energy provided at room temperature is about $25~ meV$. For large holes this energy difference can reach values around $150$meV. Thus, higher temperatures are needed to increase the healing probability. Additionally, we also generated energy mappings using the SCC-DFTB method \cite{DFTB,SCCDFTB}, in order to contrast with the data obtained from ReaxFF simulations. We considered the case of a hole with radius of $3.2$~\AA~  and a carbon probe atom placed at a distance of $3.2$ \AA~  from the membrane. As expected, although the intrinsic barrier values are different ($30$ and $80$~ meV for ReaxFF and SCC-DFTB, respectively), the general trends of the maps are qualitatively similar. A figure of the energy mapping using these two methods can be found in the supplementary material \cite{supplementary_material}.

\subsection{Temperature Dependence}
 
In order to investigate the temperature dependence on the graphene healing mechanism, we performed molecular dynamics simulations in the temperature range of $300~$K up to $2000~$K. The model system consists of a graphene membrane in which a hole is etched in its center. Also, carbon atoms are added at regular time intervals to mimic the presence of a carbon source (see methodology section for a complete description). The filling of holes is expected to depend mainly on the presence of thermal fluctuations that should be high enough to allow the added atoms to cross the energy barrier and enter into the defective region, in addition to the thermal fluctuations of the atoms from the defect itself.
\begin{figure}[h]
\centering
{\includegraphics[width=0.65\linewidth]{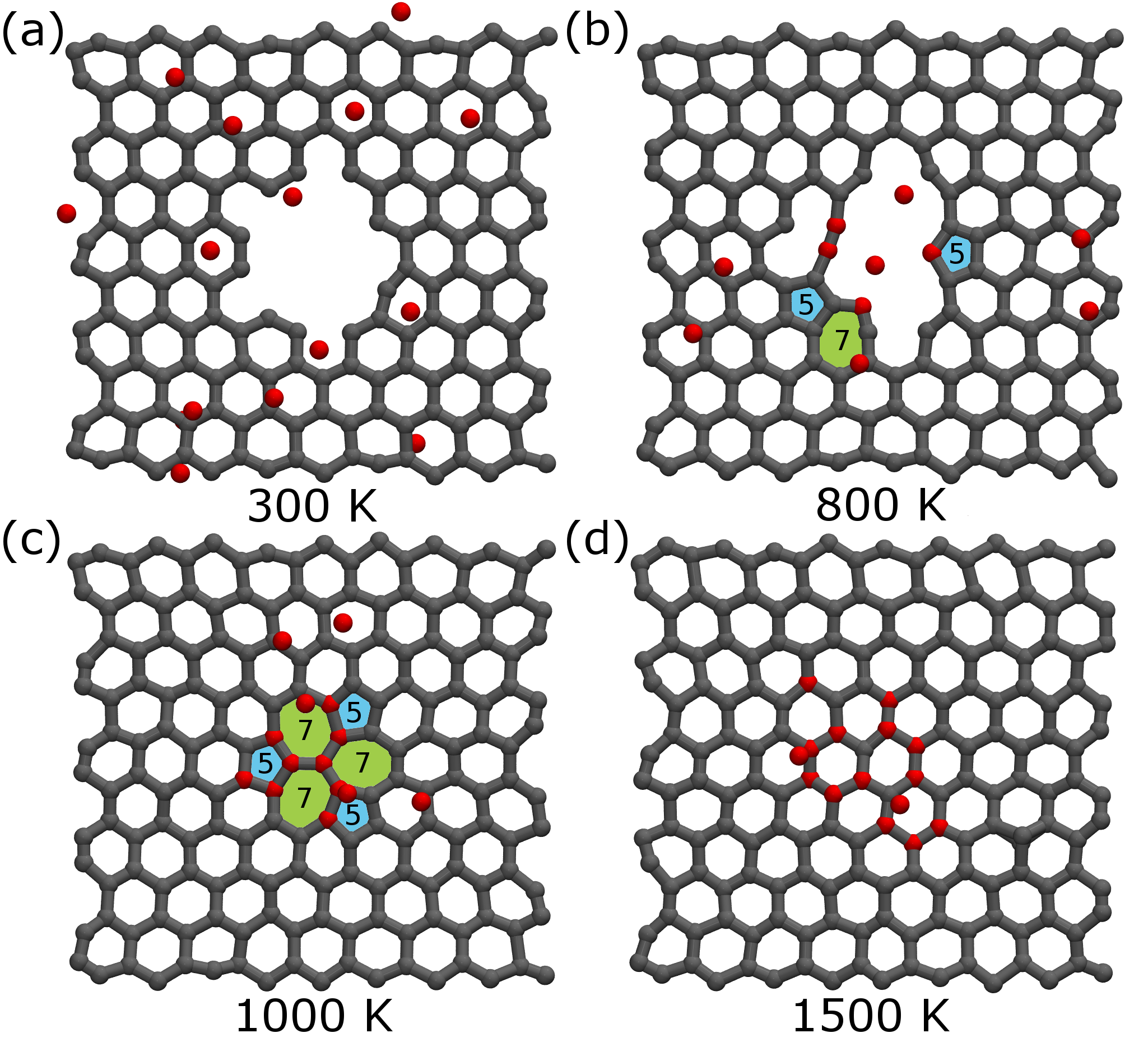}}\quad
\caption{Final obtained structures from molecular dynamics simulations at different temperatures: (a) $T=300~$K; (b) $T=800~$K; (c) $T=1000~$K; (d) $T=1500~$K. The added carbon atoms are indicated in red. The formation of rings containing 5 and 7 atoms are highlighted.}  
\label{figure3}
\end{figure}

	For $300~$K, the added atoms do not have the necessary energy to reach the chemically active region, located at the edges of the defect, and no complete filling was observed, thus confirming the trends obtained from the energy map analysis. The obtained final structure is shown in Figure \ref{figure3} (a).
 Increasing the temperature enhances the reactivity of the added atoms and of the graphene atoms near the hole region, but the healing is also incomplete for temperatures up to $800~$K. In this range of temperature, some linear atomic chains (LACs) can be found, but their inter-conversion into ring structures was not observed, see Figure \ref{figure3}(b). In addition, we observed some reconstruction of hole edges. For $1000~$K, the thermal fluctuation reaches a point in which the energy is sufficiently large to convert these LACs into more stable structures (5 up to 8 member rings) and an imperfect healing takes place, as shown in figure \ref{figure3}(c). For $1500~$K, a perfect healing was observed, since the thermal fluctuations allow the conversion of non-hexagonal to hexagonal rings, as shown in Figure \ref{figure3}(d). The graphene healing induced by annealing at high temperature was obtained by different experimental investigations, for instance, for temperatures of $800~^{o}C$ ($1073.15~$K) \cite{Niu2013} and $900~^o$C ($1173.15~$K) \cite{Chen2013,Kholmanov2011}, in good agreement our results.

\subsection{Electron Beam Effects}

To analyze the healing mechanism in similar conditions to those reported by Zan {\it et al.} \cite{Zan2012}, we have investigated other possibilities that could trigger hole filling mechanisms. Zan {\it et al.} \cite{Zan2012} observed the healing of large holes in graphene at room temperature during low energy STEM observations. They suggest that the STEM electron beam could act as a local heating mechanism, thus allowing the carbon atoms from impurities to come closer to the hole and to fill it.

\begin{figure*}[top]
\centering
\includegraphics[width=0.93\linewidth]{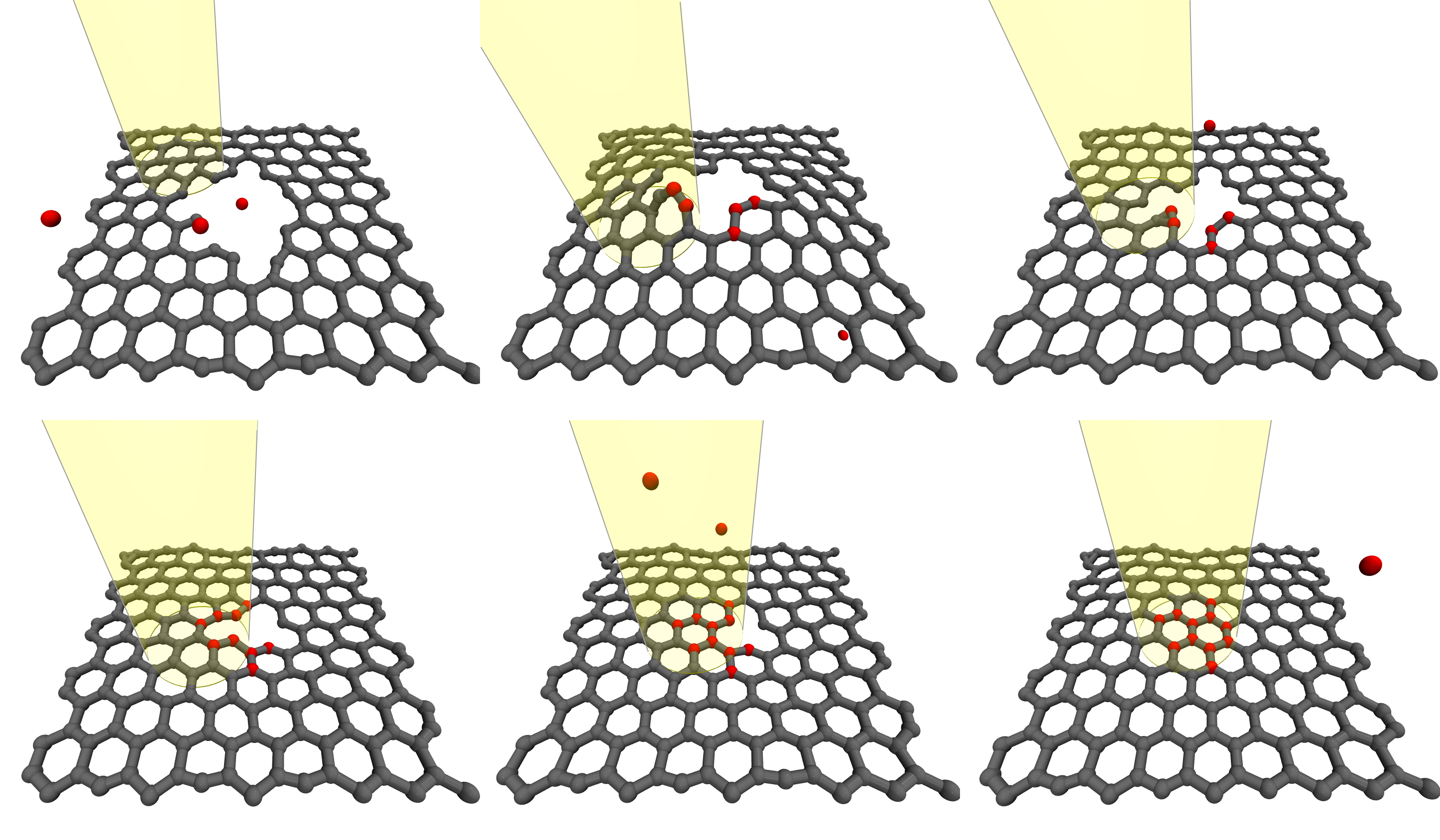}
\caption{Sequence of different configurations from MD simulations leading to a perfect healing of a defective graphene membrane at room temperature. Local heating areas (for the case of energy rate of  $0.5~kcal/(mol.fs)$), are indicated in yellow color. Graphene atoms and added carbon ones are displayed in grey and red colors, respectively.}
\label{figure4}
\end{figure*}

\begin{figure}[top]
\centering
\includegraphics[width=0.4\linewidth]{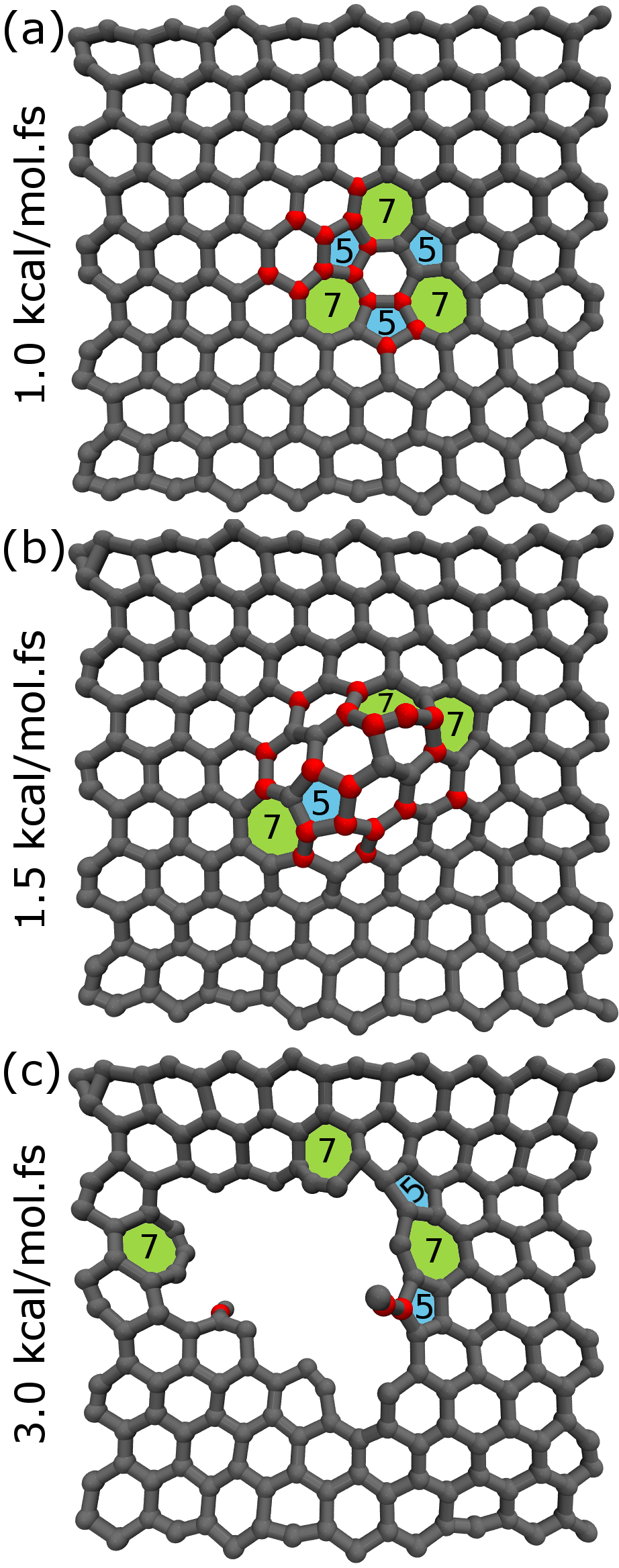}
\caption{Snapshots from MD simulations showing  the final structures for imperfect healing of a hole in a graphene membrane: (a) local heating of $1.0~kcal/(mol.fs)$, resulting in the formation of a flat defect (FD) structure; (b) local heating of $1.5~kcal/(mol.fs)$, resulting in the formation of a Lump Defect (LD) structure; (c) local heating of  $3.0~kcal/(mol.fs)$, resulting in a larger hole. See text for discussions.}
\label{figure5}
\end{figure}

\begin{figure}[top]
\centering
\includegraphics[width=1.0\linewidth]{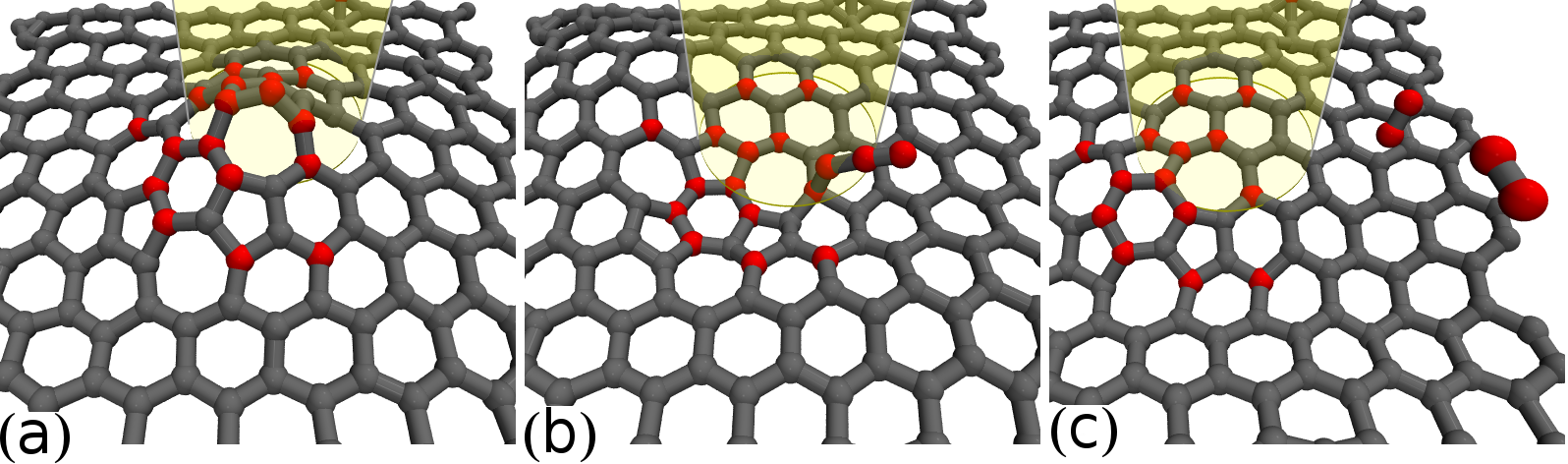}
\caption{Structural transformation in a defective graphene membrane driven by local heating (indicated in yellow color) of $2.0~kcal/(mol.fs)$: (a) Lump defect generated from high energy flux; (b) Lump defect interconversion to a LAC (linear atomic chain) structure; (c) LAC structure detachment from the defective graphene.}
\label{figure6}
\end{figure}

The mechanisms of electron beam interactions and manipulations of nanostructures are well-know \cite{Krasheninnikov2007,Krasheninnikov2010,Banhart,Petkov2013}. Banhart \cite{Banhart} emphasizes that when energetic electrons or ions strike a target, different mechanisms of energy or momentum transfer can take place. The main contributions for radiation effects are due to, amongst others: electronic excitations or ionization of individual atoms, creation of collective electronic excitations, bond breaking, phonon generation and atom displacements in the interior of the sample \cite{Banhart}.
The secondary effects are emission of photons and emission of Auger electrons \cite{Banhart}. Moreover, electron beams of low energy present a more intense interaction with the target \cite{Krasheninnikov2007}, thus transferring energy more efficiently to the lattice than a higher energy beam.
Thus, elastic collisions and some mechanisms pointed out by Banhart \cite{Banhart} could constitute the main contributions for heating of the target. 

In order to mimic these experimental conditions, we incorporated into the simulations the main effects of the interaction between the electron beam and the target, by using a local heating source in a cylindrical region of space, as shown in figure \ref{figure1}. This local heating provides energy at a constant rate to a specific spot. Also, this spot moves through the membrane mimicking the movement of the electron beam scanning in the STEM experiments. With this simulation protocol, it is possible to investigate different conditions of the electron beam by controlling the energy rate values.
The local heating approach can be useful in tailoring complex structures 
or even to sculpt free-standing graphene \cite{Schneider2013}. In some cases, applying high temperature through the whole system could cause the destruction of the complete structure, which can be prevented by the use of heating spots. For this kind of situation, a local heating source as the one we are using here, can work as a precision tool for fixing local defects or even to produce local chemical modifications.

Here, we have analyzed the effect of a simulated heating source with an energy rate in the range of $0$ to $3.0~kcal/(mol.fs)$, in the presence of a thermostat set to maintain a fixed temperature of $300~$K on atoms located outside the heated region. For energy rates above $3.0~kcal/(mol.fs)$, we observed structural damages with an increase of the hole size. 

For a local heating spot with energy rates up to $0.25~ kcal/(mol.fs)$ ($10.08~ meV/fs$), we observed no hole filling. Added atoms did not receive the necessary amount of energy to overcome the energy barriers in the interface between the membrane and the defective region. However, an energy rate of $0.5~ kcal/(mol.fs)$ ($21.68~ meV/fs$) can lead to a perfect healing. Some typical simulation snapshots of this regime are presented in figure \ref{figure4} (the entire simulation video can be seen in supplementary material \cite{supplementary_material}). In this case, the added atoms acquire the necessary energy to approach the hole edge and to react with it, filling and reconstructing the hole. With this energy rate, the local heat was able to provide the necessary amount of kinetic energy to facilitate the absorption of added atoms. Another effect of this local heating is the increase of local chemical reactivity at the hole edges, caused by out-of-plane thermal fluctuations of atoms in that region. When the hole is completely filled, new coming atoms are deflected (bounced off) by the healed membrane.

Increasing the energy rate of local heating above $0.5 ~kcal/(mol.fs)$ is likely to be effective in filling the hole, but it could also generate structures that are not perfectly hexagonal. We also observed the incorporation of more atoms than necessary to obtain perfect healing, forming a defective re-knit structure.
We named these defective structures Flat (FD) and Lump (LD) Defects. FD structures consist of planar or quasi-planar structures containing non-hexagonal rings (5, 7 and 8 atom ring), while the LD defects are those which deviate from the membrane plane rendering the ``lump'' membrane with a 3D structure similar to that proposed by Lusk and Carr \cite{Lusk2008}. FD and LD structures can be seen in Figure \ref{figure5}(a) and Figure \ref{figure5}(b), respectively. 

For an energy rate of $1.0 ~kcal/(mol.fs)$ ($43.36~ meV/fs$), the local heating is more likely to generate FD structures observed in the final healing processes, as shown in figure \ref{figure5} (a). LD structures appear with the increase of the local heating energy rate, for instance, for $1.5 ~kcal/(mol.fs)$ ($65.05~ meV/fs$), as shown in figure \ref{figure5} (b). For this energy value, the local heating eases the incorporation of added atoms causing some of them to be permanently incorporated as part of the healed structure.

\begin{figure*}[top]
\centering
\includegraphics[width=1.0\linewidth]{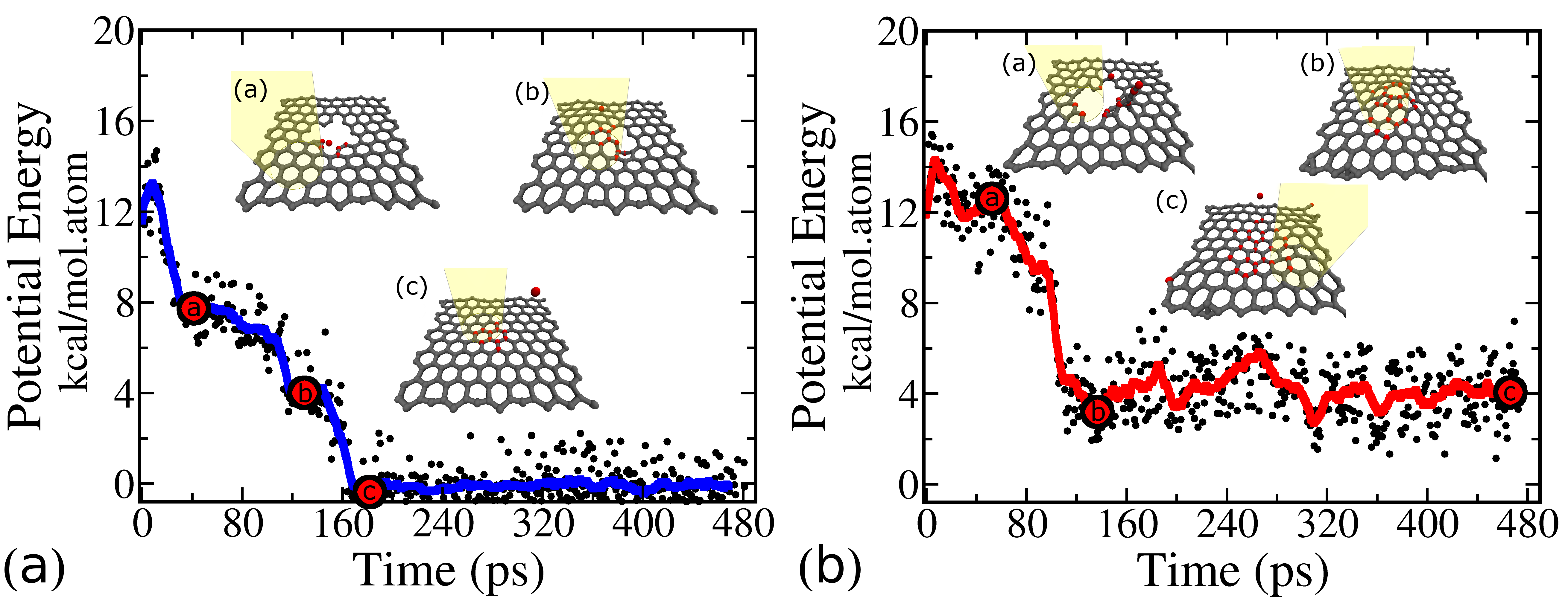}
\caption{Potential energy for different energy rates during the simulation of a healing process. Red dots represent snapshots shown in the insets a,b and c: (a) $0.5~kcal/(mol.fs)$ local heating; (b) $2.5~kcal/(mol.fs)$ local heating.}
\label{figure7}
\end{figure*}

Once they are healed, FD and LD structures remain in the membrane to energy rates up to $1.5 ~kcal/(mol.fs)$. Up to this energy rate, the transferred energy does not affect the regular bonded atoms from stable structures,{\it i.e.}, hexagonal, pentagonal and heptagonal rings. For energy rates around $2.0~ kcal/(mol.fs)$ ($86.73~ meV/fs$), it was possible to observe structural reconstructions {\it via} rotating bonds, atom migrations and also atom losses. A competition between addition and removal of carbon atoms occurs and LD structures are converted into more 2D-like structures. Snapshots of a process in which a LD structure is converted into a FD one, expelling two $C_2$ molecules, is shown in figure \ref{figure6}.

	For a $2.5~ kcal/(mol.fs)$ ($108.41~ meV/fs$) energy rate for the local heating, the competition between addition and removal of atoms reaches an interesting point: perfect healing occurs again but with different aspects than those found for $0.5 ~kcal/(mol.fs)$. A competition between formation and destruction of defective rings in the structure, edge reconstruction, bonded atom diffusion and other mechanisms are observed. The LD structures in those conditions are rapidly converted into FD structures by atom removals. The final structure obtained after the local heating scanning of the membrane can be a perfectly healed hexagonal structure. The potential energy values of intermediate structures are shown in figure \ref{figure7} (a) and (b), for $0.5 ~kcal/(mol.fs)$ and $2.5 ~kcal/(mol.fs)$, respectively. Similar plots for intermediate structures and the complete movie of the process for other values of heat rate are included in the supplementary material \cite{supplementary_material}.

When the local heating has an energy rate of $3.0 ~kcal/(mol.fs)$ ($130.09 ~meV/fs$), the filling of a hole does not occur anymore, and an etching process begins to increase the size of the hole, 
as shown in Figure \ref{figure6}(c).  The minimum energy in STEM microscopy needed to remove an atom, {\it i.e.}, to produce the knock-on effect, from a nanotube section perpendicular to the electron beam is equal to $86~ keV$ \cite{Zobelli2007} and for pristine graphene about $80~ keV$ \cite{Meyer2012}, a value higher than that used by Zan {\it et al.} \cite{Zan2012} ($60~keV$). Some experimental works of graphene exposed to high-energy electron beam show an increase on the number of vacancies under these conditions \cite{Xu2012}, which is in qualitative agreement with our results for this regime of local heating.

\section{Conclusions}

We have investigated the healing mechanisms of graphene membranes containing large holes. We have considered different experimental conditions reported in the literature through fully atomistic reactive molecular dynamics simulations: (i) the healing induced by high temperatures \cite{Niu2013,Chen2013,Kholmanov2011} and; (ii) heating effects caused by electron beam irradiation in STEM experiments leading to room temperature self-healing mechanisms\cite{Zan2012, Song2011}.

The creation of a hole in graphene structures generates significant modifications in the energy landscape experienced by atoms near the membrane. Our calculations indicate that a carbon probe atom placed near the surface should overcome an energetic barrier, as great as $150~meV$, in order to reach the interior of the defect and to initiate the healing process. Thus, depending on the energy of these atoms, the hole cannot be effectively filled. However, after the atoms are trapped inside the hole, the presence of this barrier can confine them, favoring their reaction with the membrane edges and contributing to the healing process.

In the case (i) of healing induced by temperature, we demonstrate that the kinetic energy available at room temperature is not enough to allow free atoms to overcome the energetic barriers and trigger the healing process. On the other hand, increasing the temperature ($\sim 1000~$K) could enhance dramatically the probability of an adatom to overcome the energy barrier and interact with atoms at the defect edges. Moreover, thermal fluctuations of atoms located at the hole edges can contribute to increased local reactivity, allowing reconstructions which can result in relatively stable structures (5 up to 8 member rings). Perfect healing of defective membranes was observed at temperatures of $\sim 1500$K in our simulations, which are in good agreement with experimental results \cite{Chen2013,Kholmanov2011,Xu2012,Cheng2012}.

Considering the case (ii) of STEM experiments, in which the self-healing mechanism at room temperature \cite{Zan2012} and electron-beam sculpting \cite{Song2011} were reported, we mimicked the heat effects of STEM experiments by the introduction of a local heating protocol. Using this protocol, different processes can occur depending on the applied energy rate. In one extreme, low energy rates (less than $0.25~kcal/mol.fs$) are not able to trigger the process while, in the other extreme, energy excess (from $3.0~kcal/mol.fs$ in our simulations) can lead to an increase of the defects as reported in some experimental  conditions \cite{Xu2012}. For the case of intermediate energy rates we were able to demonstrate the possibility of healing resulting in perfect hexagonal structures, as demonstrated for $0.5~kcal/(mol.fs)$ and $2.5~kcal/(mol.fs)$ energy rates. Although these two cases lead to perfect hexagonal healing, their dynamics are quite different. For the former, the process is very soft, the added atom receives only the energy required to make it enter into the defect region and to fill it, while in the latter, fast dynamics take place, with a competition between reconstruction and destruction of structures. 

We have addressed the different healing mechanisms showing the different aspects between graphene healing at high temperature and self-healing mechanism at room temperature in the presence of an electron-beam. Our conclusions about the conversion of defective structures and potential energy landscape can be extended to the recent experimental observations \cite{Zhao2014,Robertson2013} regarding adatoms interacting with graphene defects.

\section*{Suplemmentary Material}

Molecular dynamic simulations of the perfect healing mechanism in three different conditions are available in video: (i) at high temperature (1500K); at room temperature for a local heating of (ii) $0.5~ kcal/(mol.fs)$ and (iii) $2.5~ kcal/(mol.fs)$. Complementary calculations are available in a PDF document.

\section*{Acknowledgement}

This work was supported in part by the Brazilian Agencies, CAPES, CNPq and  FAPESP. The authors thanks D. Ugarte, V.T. Santana and D. Hicks for fruitful discussions. This research was supported in part by resources supplied by the Center for Scientific Computing (NCC/GridUNESP) of the S\~ao Paulo State University (UNESP). The authors thank the Center for Computational Engineering and Sciences at UNICAMP for financial support through FAPESP/CEPID grant n.2013/08293-7.

\bibliography{bib} 
\bibliographystyle{elsarticle-num}

\end{document}